\documentclass[preprint,prc,showpacs,preprintnumbers,amsmath,amssymb,floatfix]{revtex4}
\usepackage{graphicx,latexsym,amssymb, epsfig}
\usepackage{multirow,amsmath,array,booktabs}
\usepackage{subfigure}
\usepackage{color}
\usepackage{bm}
\usepackage[figuresright]{rotating}
\usepackage[section]{placeins}
\usepackage{slashed}
\usepackage{graphicx}
\usepackage{tikz}

\newcommand{\nc}{\newcommand}       
\nc{\vc}[1] {\mbox{\boldmath $#1$}} 
\nc{\del}       {\partial}              
\nc{\bra}       {\langle}               
\nc{\ket}       {\rangle}               
\nc{\bras}[1]   {\langle #1|}           
\nc{\kets}[1]   {|#1\rangle}            
\nc{\mapleft}[1]{           
 \smash{\mathop{\,          %
  \hbox to 1.5cm{\rightarrowfill}\, }\limits_{#1}}}
\nc{\beq}     {\begin{eqnarray}} \nc{\eeq}    {\end{eqnarray}}
\nc{\nn}      {\\\nonumber} \nc{\vs}      {\vspace{-0.275cm}}
\nc{\fra}    {\frac{1}{2}}
\nc{\mb}        {\mathbf}

\begin{document}

\title{Bayesian truncation errors in equations of state of nuclear matter with chiral nucleon-nucleon potentials}

\author{Jinniu Hu}
\email{hujinniu@nankai.edu.cn}
\affiliation{School of Physics, Nankai University, Tianjin 300071,  China}
\author{Peiyu Wei}
\affiliation{School of Physics, Nankai University, Tianjin 300071,  China}
\author{Ying Zhang}
\affiliation{Department of Physics, Faculty of Science, Tianjin University, Tianjin 300072, China}


\date{\today}
\begin{abstract}
The truncation errors in equations of state (EOSs) of nuclear matter derived from the chiral nucleon-nucleon ($NN$) potentials at different expansion orders are analyzed by a Bayesian model. These EOSs are expanded as functions of a dimensionless parameter, $Q$, which is determined by Fermi momentum, $k_F$ and breakdown scale, $\Lambda_b$. The degree-of-belief (DoB) intervals predicted by the chiral effective field theory are calculated within the corresponding expansion coefficients and the specific prior probability distribution functions in terms of Bayes’ theorem.  The truncation errors of EOSs, generated by the DoB intervals, exhibit good order-by-order convergences with different chiral expansion order potentials. When DoB  is considered as $1\sigma$ credibility, i.e., $68.27\%$ confidence interval, the truncation errors of binding energy per nucleon in symmetric nuclear matter and pure neutron matter are consistent with the results given by a simple error analysis method proposed by Epelbaum, {\it et al.} Finally, the reasonable values of the breakdown scale $\Lambda_b$ in nuclear matter are discussed through consistency checks of Bayesian method based on the calculations of success rate.
\end{abstract}



\maketitle

\section{Introduction}
The chiral effective field theory (ChEFT) as a perturbation method, achieves a great success in investigating the properties of hadron systems, which partially solved the difficulty of the non-nonperturbative of QCD theory at low-energy scale. Those physical quantities, e. g., nucleon mass, pion decay constant, and low energy constants were calculated from leading order to higher orders in ChEFT~\cite{scherer02,bernard06,holt16}. Furthermore, the nucleon-nucleon ($NN$) potentials, which play a very important role in nuclear physics, can be also produced from ChEFT, proposed by Weinberg firstly~\cite{weinberg90,weinberg91,weinberg92}. With the significant progresses in last thirty years~\cite{epelbaum09,machleidt11}, the chiral potentials up to next-to-next-to-next-to-next-to-leading order (N$^4$LO) were worked out by several groups
recently~\cite{epelbaum15a,epelbaum15b,entem15,entem17,reinert18}. These $NN$ potentials can provide very high-precision theoretical descriptions to the thousands of $NN$  scattering data. They have been used to study the nuclear system from light nuclei~\cite{binder16} to infinite nuclear matter~\cite{hu17}. Furthermore, the many-nucleon forces can be also obtained systematically in the unified framework of ChEFT~\cite{ishikawa07,bernard07,bernard11,epelbaum06,epelbaum07}.

The pivotal technology in ChEFT is how to take the power counting for a dimensionless parameter in the process of chiral expansion, which should ensure the higher-order contributions always smaller than the lower ones. The matrix elements of available chiral potentials certainly satisfy such requirement~\cite{machleidt11}.  However, the convergence of calculation results in the complex nuclear systems generated by chiral potentials at different expansion orders are not so obvious due to the complication of few-body and many-body methods.

To estimate the truncation errors of the chiral expansion for the nuclear observables, Epelbaum {\it et al.} developed an easily operational analysis methodology~\cite{epelbaum15a,epelbaum15b}. It was assumed that the observable can be expanded as a polynomial up to the highest chiral expansion order, whose expansion parameter $Q\in\{p/\Lambda_b,~m_\pi/
\Lambda_b\}$ is related to the typical momentum scale $p$, pion mass $m_\pi$, and breakdown scale of the chiral expansion, $\Lambda_b$. Then, the truncation errors are determined by the expansion coefficient, which has the maximum magnitude. In this way, truncation errors of many nuclear observables calculated by the latest chiral potentials from Epelbaum {\it et al.} (named as EKM potentials)~\cite{epelbaum15a,epelbaum15b}, such as the proton-neutron total cross sections, the tensor analyzing powers in elastic nucleon-deuteron scattering, the ground-state properties of light nuclei, and so on, were obtained and demonstrated the valid convergence of the chiral expansion for different order potentials~\cite{epelbaum15a,epelbaum15b,reinert18,binder16}. Recently, we also applied this scheme to investigate the truncation errors of basic properties of nuclear matter~\cite{hu17}, whose equations of state were calculated in the framework of Brueckner-Hartree-Fock (BHF) method and with EKM potentials. The valid convergence of chiral potentials for nuclear matter was affirmed.

In the meantime, a Bayesian approach was introduced to discuss the truncation errors of nuclear observable from ChEFT by Furnstahl {\it et al.}~\cite{furnstahl15,wesolowski16,melendez17}. These formalism for truncation errors were firstly proposed to investigate the quantities of perturbative QCD theory by Carriari and Houdeau~\cite{cacciari11,bagnaschi15}. This scheme was also applied to study the theoretical uncertainties from the effective filed theory for nuclear vibration by P\'erez and Papenbrock~\cite{perez15}.   In such method, the truncation errors were seriously interpreted by statistical technology. It can be derived from the degree-of-belief (DoB) interval, which is determined by the prior probability distributions from the expansion coefficients. Various $NN$ scattering observables were discussed in this framework~~\cite{furnstahl15,melendez17}. They found that the truncation errors of phase shifts with EKM potentials for $1\sigma$ DoB intervals were consistent with those from the simple analysis method proposed by Epelbaum {\it et al.}~\cite{epelbaum15a,epelbaum15b}. The statistical consistency of the chiral breakdown scale, $\Lambda_b$, was also explored.

In this work, we would like to extend this Bayesian method to analyze the truncation error of equations of state of nuclear matter obtained by BHF method and EKM potentials for different DoB intervals and compare them with our previous results obtained by the simple analysis method~\cite{hu17}. Furthermore, the reasonable breakdown scale in nuclear matter will be discussed. This paper is organized as follows. In section II, we simply show the necessary formula in Bayesian method for truncation error analysis. The results of this work are presented and discussed in section III, while conclusions are given in section IV.

\section{Bayesian truncation error analysis method}
We followed the Bayesian scheme to treat the truncation errors given by Furnstahl {\it et al.}~\cite{furnstahl15,melendez17}.  A nuclear observable $X$ in ChEFT can be expanded as a polynomial with a dimensionless parameter $Q$, which is relevant to the momentum and breakdown scale,
\beq\label{epeq}
X=X_\text{ref}\sum^\infty_{n=0} c_nQ^n,
\eeq 
where $X_\text{ref}$ is defined as the natural size of $X$. It is usually taken the leading-order value of $X$, i.e., $X_0$. $c_n$ is a dimensionless expansion coefficient and $c_1=0$ due to the symmetry requirement. In this work, we concentrate on the truncation errors of the equations of state (EOSs) of nuclear matter. Therefore, the observable is the energy per particle of nuclear matter, $E/A$, and the expansion parameter is regarded as $Q=k_F/\Lambda_b$. $k_F$ is the Fermi momentum of nucleon, which is determined by the nuclear density, $\rho_B$. $\Lambda_b$ is the ChEFT breakdown scale.

The error of observable truncating at the order $k$ of the expansion is defined as $X_\text{ref}\Delta_k$, where the dimensionless function, $\Delta_k$ is given as,
\beq
\Delta_k=\sum^\infty_{n=k+1} c_nQ^n.
\eeq
In practice, this summation for $n$ should be stopped at some $h+k+1$ order and the higher-order terms will be neglected. However, these coefficients $c_n~(n\ge k+1)$ cannot be obtained from the relevant nuclear many-body method, which should be extracted by the known expansion coefficients, $c_n~(n\le k)$.   In Bayesian model, a probability distribution function (pdf) for $\Delta_k$ is defined as $\text{pr}_h(\Delta|\bm{c_k})$, which is determined by a vector composed of the lower-order coefficients, $\bm{c_k}$. Here, $h$ means that only $h$ higher-terms are included in the truncation error. The $c_0$ and $c_1$ are excluded in vector $\bm{c_k}$, since $c_0$ is dependent on the natural size of $X$ and $c_1=0$ due to the symmetry requirement in ChEFT,
\beq
\bm{c_k}\in\{c_2,~c_3,~\dots,~c_k\}.
\eeq

This pdf decides the degree-of-belief (DoB), $p$, with the highest posterior density (HPD),
\beq\label{dobe}
p=\int^{d^{(p)}_k}_{-d^{(p)}_k}\text{pr}_h(\Delta|\bm{c_k})d\Delta,
\eeq 
where the $(100\times p)\%$ is the probability for the true value  of the nuclear observable $X$ staying in $\pm X_\text{ref}d^{(p)}_k$ at the $(k+1)$ order (N$^k$LO) prediction. 

With a Bayesian network, Furnstahl {\it et al} derived $\Delta_k$ in terms of the expansion coefficients by assuming them as random variables, which own a shared distribution with a specific size or upper bound $\bar c$ at central region. The pdf function, $\text{pr}_h(\Delta|\bm{c_k})$, generated by Bayesian theorem, finally can be written as 
\beq
\text{pr}_h(\Delta|\bm{c_k})=\frac{\int^\infty_0d\bar c \text{pr}_h(\Delta|\bar c)\text{pr}(\bar c)\prod^k_{n=2}\text{pr}(c_n|\bar c)}{\int^\infty_0d\bar c\text{pr}(\bar c)\prod^k_{n=2}\text{pr}(c_n|\bar c)}.
\eeq 
The choices of priors $\text{pr}(c_n|\bar c)$ and $\text{pr}(\bar c)$ are very important to determine $\Delta_k$. Here, we take,
\beq\label{proch}
\text{pr}(c_n|\bar c)&=&\frac{1}{2\bar c}\theta(\bar c-|c_n|),\nn
\text{pr}(\bar c)&=&\frac{1}{\sqrt{2\pi}\bar c\sigma}e^{-(\ln \bar c)^2/2\delta^2}.
\eeq 
A log-normal distribution is used with a hyperparameter $\delta$ and is normalized for $\bar c$ in $(0,~\infty)$, which was demonstrated to be consistent with other choices as shown in Refs.~\cite{furnstahl15,melendez17} for $NN$ scattering.  Therefore, in this work, we concentrate to discuss the truncation errors of nuclear matter with these choices.  Furthermore, the prior $\text{pr}_h(\Delta|\bar c)$ can be worked out with the properties of step function,
\beq
\text{pr}_h(\Delta|\bar c)=\frac{1}{2\pi}\int^\infty_{-\infty}dt \cos(\Delta t)\prod^{k+h}_{i=k+1}\frac{\sin(\bar c Q^it)}{\bar c Q^it}.
\eeq 
With these equations, the DoB interval, $d^{(p)}_k$ can be solved numerically in Eq.~(\ref{dobe}) as an inversion problem. 

\section{Numerical results and discussions}
The binding energy per nucleon of symmetric nuclear matter (SNM) and pure neutron matter (PNM) has been calculated order-by-order in the framework of Brueckner-Hartree-Fock (BHF) method with the latest chiral $NN$ potential regularized by a coordinate semilocal form factor~\cite{hu17}. In BHF model, the three-body force is very important to obtain the reasonable saturation properties. In this discussion, the analysis of truncation error for chiral potentials in nuclear matter is based on the two-body forces, which should generate the dominant contribution to the binding energies of nuclear many-body systems, while the three-body force is expected to provide the smaller contribution in the nuclear problem at low energy scale. Actually, the description accuracy of nuclear few-body system with {\it ab initio} methods will be largely improved with three-body force~\cite{epelbaum15a,epelbaum15b}. 

In the present BHF model, the Brueckner two-hole-line expansion is used. Recently, three-hole-line contribution of Goldstone diagrams was worked out, whose magnitude is less than $8\%$ of the one from two-hole-line expansion \cite{lu17}. Therefore, the higher-order many-body contributions from BHF method provide very few effects in the truncation error analysis for chiral potentials in nuclear matter.

Firstly, the expansion coefficients $c_n$ from NLO to N$^4$LO in Eq.~(\ref{epeq}) for the binding energy per nucleon are extracted. Here, the chiral potentials are adopted with cutoff $R=1.0$ fm in the semilocal form factor. The equations of state generated by another cutoff potentials have the similar behaviors.  Actually, the breakdown scale $\Lambda_b$ is a very important quantity, which determines the magnitude of the dimensionless parameter $Q$ and the expansion coefficients, $c_n$. In this work, we firstly follow the choice by EKM as $\Lambda_b=600$ MeV generated from the analysis of $NN$ scattering data~\cite{epelbaum15b}. In the last part of this section, it will be discussed whether this value is suitable for the nuclear matter system.

Here, the equations of state are calculated up to $\rho_B=0.4$ fm$^{-3}$. The corresponding Fermi momentum is about $350$ MeV for SNM, which is not so large, comparing to the conventional breakdown scale, $\Lambda_b=600$ MeV. Therefore, the chiral expansion is validated up to such density. On the other hand, the contribution of three-body force will increase rapidly with density. However, at each chiral expansion order, their magnitudes are similar, which were shown in recent work by Sammarruca {\it et al}~\cite{sammarruca18}. Therefore, the contributions of many-body interactions can reach the convergence order by order.

\begin{figure}[!hbt]
	\includegraphics[width=0.4\textwidth]{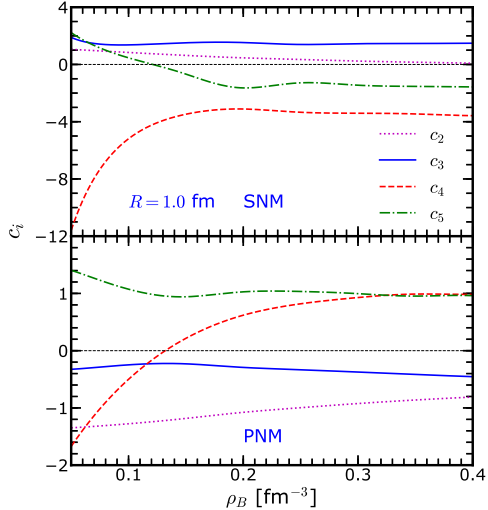}
	\caption{(Color online) The expansion coefficients, $c_n$ of binding energy per nucleon of symmetry nuclear matter (upper panel) and pure neutron matter (lower panel) as a polynomial form within different chiral potentials, whose cutoffs are $R=1.0$ fm.}\label{cn}
\end{figure}

The extracted coefficients ($c_2,~c_3,~c_4,~c_5$) are shown as functions of nucleon density in Fig.~\ref{cn}. In SNM case, the magnitude of $c_2$ is less than one, smaller than other order expansion coefficients. This is because that the binding energies per nucleon with LO and NLO chiral potentials did not have much differences. While, $c_4$ at low density are quite small due to the more repulsive components provided by N$^3$LO chiral potential compared with those from N$^2$LO. In PNM, all of these coefficient are located in the region $[-2,2]$. The magnitude of $c_3$ are smallest now. These curves of coefficients become flat at high density.

Once the expansion coefficients are obtained, the pdfs, $\text{pr}(c_n|\bar c),~\text{pr}(\bar c),~\cdots$, can be worked out naturally with Eq. (\ref{proch}). In the calculation of prior, $\text{pr}_h(\Delta|\bar c)$, the multiplication terms should be interrupted at some order. In this work, we take $h=10$, since the results are converged quickly in the present numerical methods. Furthermore, the hyperparameter, $\delta$ in prior $\text{pr}(\bar c)$ is taken as $1$ following the Ref.~\cite{furnstahl15}.  For a given DoB, $p$, like the standard confidence level, $1\sigma$ or $2\sigma$ , the integration interval, $d^{(p)}_k$ in Eq.~(\ref{dobe}) is solved inversely. Then the truncation errors at $k+1$ order of the binding energy is defined as $\pm \left(\frac{E}{A}\right)_\text{LO}d^{(p)}_k$.

\begin{figure}[!hbt]
	\includegraphics[width=0.5\textwidth]{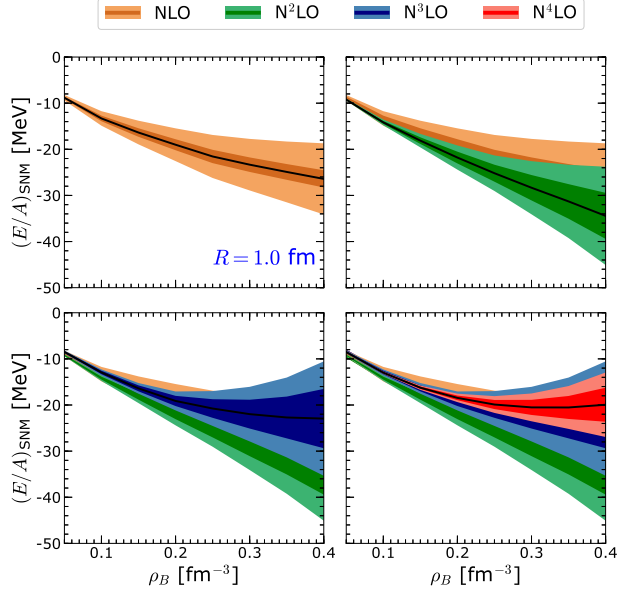}
	\caption{(Color online) The EOSs of SNM with truncation errors using chiral potentials from NLO to N$^4$LO, whose cutoffs are $R=1.0$ fm.}\label{snm}
\end{figure}

In Fig.~\ref{snm}, the EOSs of SNM with truncation errors are shown order by order. The original EOSs generated by BHF model is given as a solid curve as a function of density $\rho_B$, where $\rho_B=2k^3_F/3\pi^2$. The dark shaded bands for each color indicate the DoB interval is $1\sigma$ standard deviation, i. e., $p=68.27\%$, while the light ones corresponding to $2\sigma$ standard deviation, where $p=95.45\%$. Generally speaking, a higher DoB brings a larger uncertainty, which means the truncation errors with $1\sigma$ standard deviation is smaller than those with $2\sigma$. Therefore, the dark band should be included in the light band.

The truncation errors of binding energy per nucleon of SNM at NLO are very small, especially for the $1\sigma$ case. As we shown before, the relevant expansion coefficients, $c_2$, are quite small since the differences between the EOSs from the LO and NLO are slight. From N$^2$LO to N$^4$LO, the truncation errors decrease systematically order by order at each density. In the same chiral expansion order, they become larger with density increasing since the $Q$ value calculated by using the Fermi momentum becomes larger. This behaviors is natural since a larger $Q$ value will produce a worse expansion convergence from the perturbative point of view. The empirical saturation properties of nuclear matter ($\rho=0.16\pm0.01$ fm$^{-3}$ and $E/A=-16\pm1$ MeV) are still not satisfied after considering the Bayesian truncation errors only with two-body chiral potentials. However, it can be found that the saturation densities become smaller derived from the upper limit of error band comparing with those obtained by BHF method directly and are more closed to the empirical values.

Furthermore, it is noticed that the truncation errors with $68.27\%$ DoB given by Bayesian method are consistent with our previous results by a simpler analysis method in EKM scheme except those at NLO~\cite{hu17}.  In the EKM method, the errors of observables at NLO are estimated not only by the shifts from NLO to LO but also by LO itself. Therefore, the EKM uncertainty at NLO is much larger than those worked by Bayesian method. In the other orders, the truncations errors of $NN$ scattering observables from EKM scheme and Bayesian method also have the similar behaviors shown by Furnstahl {\it et al.} ~\cite{furnstahl15,melendez17}. 

\begin{figure}[!hbt]
	\includegraphics[width=0.5\textwidth]{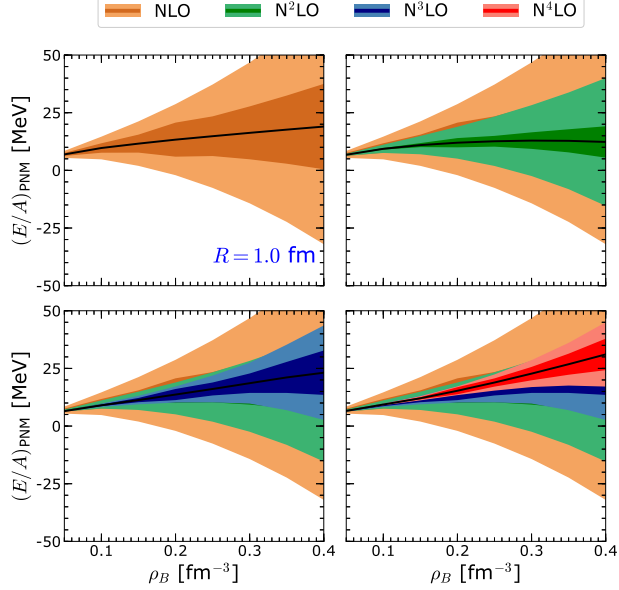}
	\caption{(Color online) The EOSs of PNM with truncation errors using chiral potentials from NLO to N$^4$LO, whose cutoffs are $R=1.0$ fm.}\label{pnm}
\end{figure}

The analogous calculations are done for the PNM. Their EOSs with uncertainties are shown in Fig.~\ref{pnm}. In the PNM, the nucleon density is given by $\rho_B=k^3_F/3\pi^2$. Therefore, with a same density, the dimensionless parameter, $Q$ in PNM is larger than that in SNM. As a result, the truncation errors of PNM are also larger comparing with the ones of SNM. The band of N$^2$LO for $68.27\%$ credibility  in PNM is relatively narrow, which is also related to the small magnitude of $c_3$ given in Fig.~\ref{cn}. In general, the uncertainty of binding energy per nucleon of PNM decreases order by order.

\begin{figure}[!hbt]
	\includegraphics[width=0.4\textwidth]{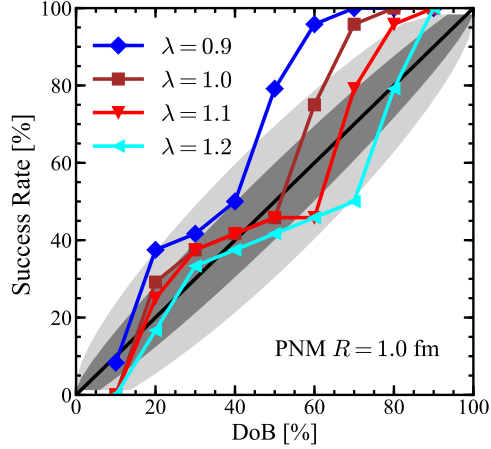}
	\caption{(Color online) Consistency check for the binding energy per nucleon of PNM using the EKM chiral potentials with $R=1.0$ fm. The ideal $1\sigma$ and $2\sigma$ confidence intervals for the success rates are shown as shaded bands.}\label{srneu}
\end{figure}

The efficacy of truncation errors analysis in this work can be examined with the Bayesian model checking. Consistency checks will be done to investigate whether the DoB intervals have been reasonably displayed as estimated. When the truncation errors from our calculations are reasonable, a fixed DoB interval $d^{(p)}_k$, at some order should include the next order contribution of the binding energy per nucleon at the same time. We will follow the procedure of consistency checked proposed by Furnstahl {\it et al.}~\cite{melendez17}. Construct a set of data ($N$ data), including the binding energies per nucleon of SNM or PNM from NLO to N$^3$LO, which has the next-order calculation results with different nuclear densities firstly. Then take a grid of DoB intervals, whose values are located in $[0,~1]$ and calculate a fixed DoB interval $d^{(p)}_k$ for binding energy per nucleon at each density in the set within the same pdfs throughout. If each next-order contribution, $c_{k+1}Q^{k+1}$ is in the DoB interval of the previous order, $c_{k+1}Q^{k+1}\in[-d^{(p)}_k,~d^{(p)}_k]$, count one success. The actual success rate is defined as the ratio of the successes $n$ to the total calculation number $N$.

In this scheme, the consistency of breakdown scale $\Lambda_b$ can be checked and its reasonable value also can be determined. In idea case, where each data in the set is uncorrelated,  the success rate should satisfy a binomial distribution. Hence, the posterior of successes with total number $N$, and  the given $p$ is,
\beq
\text{pr}(n|p,N)=\frac{N!}{n!(N-n)!}p^n(1-p)^{N-n}.
\eeq
For a continuous $n$, this posterior corresponds to a $\beta$ distribution. The confidence intervals on $n$ for a given value of DoB $p$ can be calculated with the highest posterior density prescription. In this work, we chose the $1\sigma$ and $2\sigma$ confidence intervals as ideal results.

The observable $X$ can be expanded by including a scaling factor $\lambda$ as
\beq
X=X_\text{ref}\sum_{n=0}^\infty(c_n\lambda^n)\times\left(\frac{Q}{\lambda}\right)^n.
\eeq
With this equation, the consistency checks at different $\lambda$ can be done and we can explore a more reasonable breakdown scale $\Lambda_b$ for nuclear matter.

In Fig.~\ref{srneu}, the consistency checks for the binding energy per nucleon of PNM  as an example, are displayed. The success rate of the binding energy per nucleon of PNM within NLO, N$^2$LO, and N$^3$LO chiral potentials at densities $\rho=0.05,~0.10,\cdots,~0.40$ fm$^{3}$ are calculated with the procedure mentioned before. There are $N=24$ points in this set. $\lambda=1$ corresponds to the original breakdown scale $\Lambda_b=600$ MeV.  The $\lambda$ value is taken from $0.9$ to $1.2$ in this work. The ideal success rates with $1\sigma$ and $2\sigma$ confidence intervals are plotted as gray band and light gray band, respectively. 

At $\lambda=0.9$, i.e. $\Lambda_b=540$ MeV, the success rates are higher than the limit of ideal case when DoB is larger than $50\%$. With the $\lambda$ increasing, the success rates are gradually closed to  the region of ideal case. The curves with $\lambda=1.1$ and $\lambda=1.2$ fall in the band representing $2\sigma$ confidence interval completely, which means the breakdown scale, $\Lambda_b$ should choose a more reasonable value such as $660$ MeV or $720$ MeV in the investigation of truncation error of PNM. 

The similar consistency checks for SNM are also done. It is found that the success rates of SNM with $\lambda=0.8$ are located to the idea case, which means that the breakdown scale, $\Lambda_b=480$ MeV is preferred in SNM. This is because that the expansion parameter, $Q$ in SNM is smaller than that in PNM for a fixed baryon density.

\section{Conclusion}
The Bayesian analysis method were applied to study the truncation errors in equations of state of nuclear matter generated by chiral nucleon-nucleon potentials in the framework of Brueckner-Hartree-Fock model. The binding energy per nucleon of nuclear matter was assumed to be expanded as a polynomial of dimensionless parameter, $Q$, which is dependent on the Fermi momenta and breakdown scale. The truncation errors from the chiral expansion were considered as the higher-order expansion contributions. The Bayes theorem can provide an efficient scheme to obtain the information of unknown expansion coefficients from the known ones. With this scheme, the truncation errors of binding energy per nucleon of symmetric nuclear matter and pure neutron matter were obtained with $1\sigma$ and $2\sigma$ confidence intervals at different chiral expansion orders. These truncation errors became smaller with expansion order increasing in general, which were consistent with our previous results within a simple analysis method. These calculations demonstrated that there are very good convergences for the chiral potentials in nuclear matter. Finally, the consistency checks were done to promise the reasonable analysis with success rates for different breakdown scales. They were compared to the ideal case. A larger breakdown scale in nuclear matter may be more reasonable comparing to the value from nucleon-nucleon scattering. More analysis calculations with different prior, posterior pdfs and chiral potentials will be carefully done in future. The Bayesian method provide a good chance for us to check the convergence of observables in nuclear physics worked by chiral effective field theory.

This work was supported in part by the National Natural Science Foundation of China (Grants  No. 11775119, No. 11405090, and No. 11405116).

\end{document}